\documentclass[aps,prb,twocolumn,showpacs]{revtex4}

\usepackage{graphicx}
\usepackage{amsmath}
\usepackage{amssymb}

\begin{document}

\title{Magnetic field induced incommensurate resonance in
cuprate superconductors}

\author{Jingge Zhang and Li Cheng}

\affiliation{Department of Physics, Beijing Normal University,
Beijing 100875, China}

\author{Huaiming Guo}
\affiliation{Department of Physics, Capital Normal University,
Beijing 100037, China}

\author{Shiping Feng$^{*}$}

\affiliation{Department of Physics, Beijing Normal University,
Beijing 100875, China~}

\begin{abstract}
The influence of a uniform external magnetic field on the dynamical
spin response of cuprate superconductors in the superconducting
state is studied based on the kinetic energy driven superconducting
mechanism. It is shown that the magnetic scattering around low and
intermediate energies is dramatically changed with a modest external
magnetic field. With increasing the external magnetic field,
although the incommensurate magnetic scattering from both low and
high energies is rather robust, the commensurate magnetic resonance
scattering peak is broadened. The part of the spin excitation
dispersion seems to be an hourglass-like dispersion, which breaks
down at the heavily low energy regime. The theory also predicts that
the commensurate resonance scattering at zero external magnetic
field is induced into the incommensurate resonance scattering by
applying an external magnetic field large enough.
\end{abstract}

\maketitle


\section{Introduction}

The intimate relationship between the short-range antiferromagnetic
(AF) correlation and superconductivity is one of the most striking
features of cuprate superconductors \cite{kastner}. This is followed
an experimental fact that the parent compounds of cuprate
superconductors are Mott insulators with the AF long-range order
(AFLRO) \cite{kastner}. However, when holes or electrons are doped
into these Mott insulators \cite{bednorz}, the ground state of the
systems is fundamentally altered from a Mott insulator with AFLRO to
a superconductor with persistent short-range correlations
\cite{kastner,shen}. The evidence for this closed link is provided
from the inelastic neutron scattering (INS) experiments
\cite{tranquada1,dai,yamada,bourges,he,hayden} that show the
unambiguous presence of the short-range AF correlation in cuprate
superconductors in the superconducting (SC) state.

At zero external magnetic field, the dynamical spin response of
cuprate superconductors exhibits a number of universal features
\cite{tranquada1,dai,yamada,bourges,he,hayden}, where the magnetic
excitations form an hourglass-like dispersion centered at the AF
ordering wave vector ${\bf Q}=[\pi,\pi]$ (in units of inverse
lattice constant). At the saddle point, the dispersing
incommensurate (IC) branches merge into a sharp commensurate
feature, which is dramatically enhanced upon entering the SC state
and commonly referred as the magnetic resonance scattering
\cite{tranquada1,dai,yamada,bourges,he,hayden}. In particular, it
has been argued that this commensurate magnetic resonance plays a
crucial role for the SC mechanism in cuprate superconductors, since
the commensurate magnetic resonance with the magnetic resonance
energy scales with the SC transition temperature forming a universal
plot for all cuprate superconductors \cite{wilson}. To test the
connection between the commensurate magnetic resonance phenomenon
and SC mechanism, it is desirable to perform further
characterization. Since a uniform external magnetic field can serve
as a weak perturbation helping to probe the nature of the
short-range AF correlation and superconductivity, therefore the
dynamical spin response of cuprate superconductors in the SC state
has been studied experimentally by application of a uniform external
magnetic field \cite{khaykovich,lake,tranquada,bourges1,dai1}.
However, there is no a general consensus. Some experimental results
show that applying a uniform external magnetic field enhances the
amplitude of the IC magnetic scattering already present in the
system \cite{khaykovich,lake}. On the other hand, other experiments
indicate that the intensity gain of the IC magnetic scattering is
suppressed by application of a uniform external magnetic field
\cite{tranquada}. In particular, the influence of a uniform external
magnetic field has been investigated on the resonance scattering
peak by using INS technique \cite{bourges1,dai1}. The early INS
measurement \cite{bourges1} shows that under a modest external
magnetic field ($\sim 11$ Tesla), the resonance scattering peak
remains almost unaffected, i.e., although a line broadening occurs
without change of the resonance scattering peak amplitude, no
shifting of the resonance scattering peak energy is observed.
However, the later INS experiments \cite{dai1} show that a modest
external magnetic field applied to cuprate superconductors in the SC
state yields a very significant reduction in the commensurate
magnetic resonance scattering. To the best of our knowledge, there
are no explicit microscopic predictions about the effect of a
uniform external magnetic field large enough on the magnetic
resonance scattering.

For the case of zero external magnetic field, the dynamical spin
response of cuprate superconductors has been discussed \cite{feng}
based on the framework of the kinetic energy driven SC mechanism
\cite{feng1}, and all main features of the INS experiments are
reproduced, including the doping and energy dependence of the IC
magnetic scattering at both low and high energies and commensurate
magnetic resonance at intermediate energy
\cite{tranquada1,dai,yamada,bourges,he,hayden}. In this paper, we
study the influence of a uniform external magnetic field on the
dynamical spin response of cuprate superconductors in the SC state
along with this line. We calculate explicitly the dynamical spin
structure factor of cuprate superconductors under a uniform external
magnetic field, and show that the magnetic scattering around low and
intermediate energies is dramatically changed with a modest external
magnetic field. With increasing the external magnetic field,
although the IC magnetic scattering from both low and high energies
is rather robust, the commensurate magnetic resonance scattering
peak is broadened. The part of the spin excitation dispersion seems
to be an hourglass-like dispersion, which breaks down at the heavily
low energy regime.

The rest of this paper is organized as follows. The basic formalism
is presented in Sec. II, where we generalize the calculation of the
dynamical spin structure factor from the previous zero external
magnetic field case \cite{feng} to the present case with a uniform
external magnetic field. Within this theoretical framework, we
discuss the influence of a uniform external magnetic field on the
dynamical spin response of cuprate superconductors in the SC state
in Sec. III, where we predict that the commensurate magnetic
resonance scattering at zero external magnetic field is induced into
the IC magnetic resonance scattering by an applied external magnetic
field large enough. Finally, we give a summary and discussions in
Sec. IV.

\section{Theoretical Framework}

In cuprate superconductors, the characteristic feature is the
presence of the CuO$_{2}$ plane \cite{kastner,shen}. It has been
shown from ARPES experiments that the essential physics of the doped
CuO$_{2}$ plane is properly accounted by the $t$-$J$ model on a
square lattice \cite{shen,kim}. However, for discussions of the
influence of a uniform external magnetic field on the dynamical spin
response of cuprate superconductors in the SC state, the $t$-$J$
model can be expressed by including the Zeeman term as,
\begin{eqnarray}
H&=&-t\sum_{i\hat{\eta}\sigma}C^{\dagger}_{i\sigma}
C_{i+\hat{\eta}\sigma}+t'\sum_{i\hat{\tau}\sigma}
C^{\dagger}_{i\sigma}C_{i+\hat{\tau}\sigma}+\mu\sum_{i\sigma}
C^{\dagger}_{i\sigma}C_{i\sigma}\nonumber \\
&+&J\sum_{i\hat{\eta}}{\bf S}_{i} \cdot {\bf S}_{i+\hat{\eta}}
-\varepsilon_{B}\sum_{i\sigma}\sigma C^{\dagger}_{i\sigma}
C_{i\sigma},
\end{eqnarray}
where $\hat{\eta}=\pm\hat{x},\pm \hat{y}$, $\hat{\tau}=\pm\hat{x}
\pm\hat{y}$, $C^{\dagger}_{i\sigma}$ ($C_{i\sigma}$) is the electron
creation (annihilation) operator, ${\bf S}_{i}=(S^{x}_{i},S^{y}_{i},
S^{z}_{i})$ are spin operators, $\mu$ is the chemical potential, and
$\varepsilon_{B}=g\mu_{B}B$ is the Zeeman magnetic energy, with the
Lande factor $g$, Bohr magneton $\mu_{B}$, and a uniform external
magnetic field $B$. This $t$-$J$ model with a uniform external
magnetic field is subject to an important local constraint
$\sum_{\sigma} C^{\dagger}_{i\sigma}C_{i\sigma}\leq 1$ to avoid the
double occupancy \cite{anderson}. The strong electron correlation in
the $t$-$J$ model manifests itself by this local constraint
\cite{anderson}, which can be treated properly in analytical
calculations within the charge-spin separation (CSS) fermion-spin
theory \cite{feng2,feng3}, where the constrained electron operators
are decoupled as $C_{i\uparrow}= h^{\dagger}_{i\uparrow}S^{-}_{i}$
and $C_{i\downarrow}= h^{\dagger}_{i\downarrow} S^{+}_{i}$, with the
spinful fermion operator $h_{i\sigma}=e^{-i\Phi_{i\sigma}}h_{i}$
represents the charge degree of freedom together with some effects
of spin configuration rearrangements due to the presence of the
doped hole itself (charge carrier), while the spin operator $S_{i}$
represents the spin degree of freedom (spin), then the $t$-$J$ model
with a uniform external magnetic field (1) can be expressed in this
CSS fermion-spin representation as,
\begin{eqnarray}
H&=&-t\sum_{i\hat{\eta}}(h_{i\uparrow}S^{+}_{i}
h^{\dagger}_{i+\hat{\eta}\uparrow}S^{-}_{i+\hat{\eta}}+
h_{i\downarrow}S^{-}_{i}h^{\dagger}_{i+\hat{\eta}\downarrow}
S^{+}_{i+\hat{\eta}})\nonumber \\
&+&t'\sum_{i\hat{\tau}}(h_{i\uparrow}S^{+}_{i}
h^{\dagger}_{i+\hat{\tau}\uparrow}S^{-}_{i+\hat{\tau}}+
h_{i\downarrow}S^{-}_{i}h^{\dagger}_{i+\hat{\tau}\downarrow}
S^{+}_{i+\hat{\tau}})\nonumber\\
&-&\mu\sum_{i\sigma}h^{\dagger}_{i\sigma} h_{i\sigma}+J_{\rm
eff}\sum_{i\hat{\eta}}{\bf S}_{i}\cdot {\bf S}_{i+\hat{\eta}}
-2\varepsilon_{B}\sum_{i}S^{z}_{i},~~~
\end{eqnarray}
with $J_{\rm eff}=(1-x)^{2}J$, and $x=\langle h^{\dagger}_{i\sigma}
h_{i\sigma}\rangle=\langle h^{\dagger}_{i} h_{i}\rangle$ is the hole
doping concentration. It has been shown that the electron local
constraint for the single occupancy is satisfied in analytical
calculations in this CSS fermion-spin theory \cite{feng2,feng3}.

Within the framework of the CSS fermion-spin theory
\cite{feng2,feng3}, the kinetic energy driven superconductivity has
been developed \cite{feng1}. It has been shown that the interaction
from the kinetic energy term in the $t$-$J$ model (2) is quite
strong, and can induce the d-wave charge carrier pairing state by
exchanging spin excitations in the higher power of the doping
concentration, then the d-wave electron Cooper pairs originating
from the d-wave charge carrier pairing state are due to the
charge-spin recombination, and their condensation reveals the d-wave
SC ground-state. Moreover, this SC-state is controlled by both
d-wave SC gap function and quasiparticle coherence, which leads to
that the SC transition temperature increases with increasing doping
in the underdoped regime, and reaches a maximum in the optimal
doping, then decreases in the overdoped regime \cite{feng}.
Furthermore, for the case of zero external magnetic field, the
doping and energy dependent dynamical spin response of cuprate
superconductors in the SC-state has been discussed in terms of the
collective mode in the charge carrier particle-particle channel
\cite{feng}, and the results are in qualitative agreement with the
INS experimental data on cuprate superconductors in the SC state
\cite{tranquada1,dai,yamada,bourges,he,hayden}. Following their
discussions \cite{feng}, the full spin Green's function in the
presence of a uniform external magnetic field is obtained as,
\begin{eqnarray}
D({\bf k},\omega)={1\over D^{(0)-1}({\bf k},\omega)-
\Sigma^{(s)}({\bf k},\omega)},
\end{eqnarray}
with the mean-field (MF) spin Green's function,
\begin{eqnarray}
D^{(0)}({\bf k},\omega)&=&{B_{{\bf k}}\over 2\omega_{{\bf k}}}\left
({1\over\omega-\omega^{(1)}_{{\bf k}}}-{1\over\omega+
\omega^{(2)}_{{\bf k} }}\right )\nonumber\\
&=&\sum_{\nu=1,2}(-1)^{\nu+1} {B_{{\bf k}}\over 2\omega_{{\bf k}}}
{1\over\omega- \omega^{(\nu)}_{{\bf k}}} ,
\end{eqnarray}
where $B_{{\bf k}}=2\lambda_{1}(A_{1}\gamma_{{\bf k}}-A_{2})-
\lambda_{2}(2\chi^{z}_{2}\gamma_{{\bf k}}'-\chi_{2})$, $\lambda_{1}
=2ZJ_{\rm eff}$, $\lambda_{2}=4Z\phi_{2}t'$, $\gamma_{{\bf k}}=(1/Z)
\sum_{\hat{\eta}}e^{i{\bf k}\cdot\hat{\eta}}$, $\gamma_{{\bf k}}'=
(1/Z)\sum_{\hat{\tau}}e^{i{\bf k}\cdot\hat{\tau}}$, $Z$ is the
number of the nearest neighbor or next nearest neighbor sites of a
square lattice, $A_{1}=\epsilon \chi^{z}_{1}+\chi_{1}/2$, $A_{2}=
\chi^{z}_{1}+\epsilon\chi_{1}/2$, $\epsilon=1+2t\phi_{1}/J_{\rm
eff}$, the charge carrier's particle-hole parameters $\phi_{1}=
\langle h^{\dagger}_{i\sigma}h_{i+\hat{\eta}\sigma}\rangle$ and
$\phi_{2}=\langle h^{\dagger}_{i\sigma}h_{i+\hat{\tau}\sigma}
\rangle$, and the spin correlation functions $\chi_{1}=\langle
S^{+}_{i} S^{-}_{i+\hat{\eta}}\rangle$, $\chi_{2}=\langle S^{+}_{i}
S^{-}_{i+\hat{\tau}}\rangle$, $\chi^{z}_{1}=\langle S_{i}^{z}
S_{i+\hat{\eta}}^{z}\rangle$, $\chi^{z}_{2}=\langle S_{i}^{z}
S_{i+\hat{\tau}}^{z}\rangle$, and the MF charge carrier excitation
spectrum, $\xi_{k}=Zt\chi_{1}\gamma_{{\bf k}}-Zt'\chi_{2}
\gamma'_{{\bf k}} -\mu$. Since a uniform external magnetic field is
applied to the system, the MF spin excitation spectrum has two
branches, $\omega^{(1)}_{{\bf k}}=\omega_{{\bf k}}+2\varepsilon_{B}$
and $\omega^{(2)}_{{\bf k}}=\omega_{{\bf k}}-2\varepsilon_{B}$, with
$\omega_{{\bf k}}$ is the MF spin excitation spectrum at zero
external magnetic field, and has been evaluated as \cite{feng},
\begin{eqnarray}
\omega^{2}_{{\bf k}}&=&\lambda_{1}^{2}[(A_{4}-\alpha\epsilon
\chi^{z}_{1}\gamma_{{\bf k}}-{1\over 2Z}\alpha\epsilon\chi_{1})
(1-\epsilon\gamma_{{\bf k}})\nonumber \\
&+&{1\over 2}\epsilon(A_{3}-{1\over 2} \alpha\chi^{z}_{1}-\alpha
\chi_{1}\gamma_{{\bf k}})(\epsilon-\gamma_{{\bf k}})]\nonumber \\
&+&\lambda_{2}^{2}[\alpha(\chi^{z}_{2}\gamma_{{\bf k}}'-{3\over 2Z}
\chi_{2})\gamma_{{\bf k}}'+{1\over 2}(A_{5}-{1\over 2}\alpha
\chi^{z}_{2})]\nonumber \\
&+& \lambda_{1}\lambda_{2}[\alpha\chi^{z}_{1}(1-\epsilon
\gamma_{{\bf k}})\gamma_{{\bf p}}'-{1\over 2}\alpha \epsilon(C_{3}-
\chi_{2} \gamma_{{\bf k}})\nonumber \\
&+&{1\over 2}\alpha(\chi_{1} \gamma_{{\bf k}}'-C_{3})
(\epsilon-\gamma_{{\bf p}})+\alpha \gamma_{{\bf k}}'(C^{z}_{3}
-\epsilon \chi^{z}_{2}\gamma_{{\bf k}})],~~~
\end{eqnarray}
where $A_{3}=\alpha C_{1}+(1-\alpha)/(2Z)$, $A_{4}=\alpha C^{z}_{1}+
(1-\alpha)/(4Z)$, $A_{5}=\alpha C_{2}+(1-\alpha)/(2Z)$, and the spin
correlation functions $C_{1}=(1/Z^{2})\sum_{\hat{\eta},\hat{\eta'}}
\langle S_{i+\hat{\eta}}^{+}S_{i+\hat{\eta'}}^{-}\rangle$,
$C^{z}_{1}=(1/Z^{2})\sum_{\hat{\eta},\hat{\eta'}}\langle
S_{i+\hat{\eta}}^{z}S_{i+\hat{\eta'}}^{z}\rangle$, $C_{2}=(1/Z^{2})
\sum_{\hat{\tau},\hat{\tau'}}\langle S_{i+\hat{\tau}}^{+}
S_{i+\hat{\tau'}}^{-}\rangle$, and $C_{3}=(1/Z)\sum_{\hat{\tau}}
\langle S_{i+\hat{\eta}}^{+} S_{i+\hat{\tau}}^{-}\rangle$,
$C^{z}_{3}=(1/Z) \sum_{\hat{\tau}}\langle S_{i+\hat{\eta}}^{z}
S_{i+\hat{\tau}}^{z}\rangle$. In order to satisfy the sum rule of
the correlation function $\langle S^{+}_{i}S^{-}_{i}\rangle=1/2$ in
the case without AFLRO, the important decoupling parameter $\alpha$
has been introduced in the MF calculation, which can be regarded as
the vertex correction \cite{kondo}. The spin self-energy function
$\Sigma^{(s)}({\bf k},\omega)$ in the SC-state is obtained from the
charge carrier bubble in the charge carrier particle-particle
channel as \cite{feng},
\begin{widetext}
\begin{eqnarray}
\Sigma^{(s)}({\bf k},\omega)&=&{1\over N^{2}}\sum_{{\bf p},{\bf q},
\nu=1,2}(-1)^{\nu+1} \Lambda({\bf q},{\bf p},{\bf k}) {B_{{\bf q}
+{\bf k}} \over \omega_{{\bf q}+{\bf k}}}{Z^{2}_{hF}\over 8}
{\bar{\Delta}^{(d)}_{hZ}({\bf p})\bar{\Delta}^{(d)}_{hZ}({\bf p}+
{\bf q})\over E_{\bf{p}}E_{{\bf p}+{\bf q}}}\left({F^{(\nu)}_{1}
({\bf k},{\bf p},{\bf q}\over\omega-(E_{\bf p}-E_{{\bf p}+{\bf q}}+
\omega^{(\nu)}_{{\bf q}+{\bf k}})}\right . \nonumber  \\
&+&\left . {F^{(\nu)}_{2}({\bf k},{\bf p},{\bf q})\over\omega-
(E_{{\bf p}+{\bf q}}-E_{\bf p}+\omega^{(\nu)}_{{\bf q}+{\bf k}})} +
{F^{(\nu)}_{3}({\bf k},{\bf p},{\bf q})\over\omega- (E_{{\bf p}}+
E_{{\bf p}+{\bf q}}+\omega^{(\nu)}_{{\bf q}+{\bf k}})}
-{F^{(\nu)}_{4}({\bf k},{\bf p},{\bf q})\over\omega+(E_{{\bf p}+{\bf
q}}+E_{\bf p}-\omega^{(\nu)}_{{\bf q}+{\bf k}})} \right),
\end{eqnarray}
\end{widetext}
where $\Lambda({\bf q},{\bf p},{\bf k})=(Zt\gamma_{{\bf k}-{\bf p}}-
Zt'\gamma'_{{\bf k}-{\bf p}})^{2}+(Zt\gamma_{{\bf q}+{\bf p}+{\bf k}
}-Zt'\gamma'_{{\bf q}+{\bf p}+{\bf k}})^{2}$,
$\bar{\Delta}_{hZ}({\bf k})=Z_{hF}\bar{\Delta}_{h}({\bf k})$, the
charge carrier quasiparticle spectrum $E_{h{\bf k}}=
\sqrt{\bar{\xi}^{2}_{{\bf k}}+|\bar{\Delta}_{hZ}({\bf k})|^{2}}$,
$\bar{\xi}_{\bf k}=Z_{hF}\xi_{{\bf k}}$, $\bar{\Delta}_{h}({\bf k})=
\bar{\Delta}_{h}\gamma^{(d)}_{{\bf k}}$ is the effective charge
carrier gap function in the d-wave symmetry with $\gamma^{(d)}_{{\bf
k}}=({\rm cos} k_{x}-{\rm cos}k_{y})/2$, $F^{(\nu)}_{1}({\bf k},{\bf
p},{\bf q})=n_{B}(\omega^{(\nu)}_{{\bf q}+ {\bf k}})[n_{F}(E_{\bf p}
)-n_{F}(E_{{\bf p}+{\bf q}})]- n_{F}(-E_{{\bf p}})n_{F}(E_{{\bf p}+
{\bf q}})$, $F^{(\nu)}_{2}({\bf k},{\bf p},{\bf q})=
n_{B}(\omega^{(\nu)}_{{\bf q}+ {\bf k}})[n_{F}(E_{{\bf p}+{\bf q}})-
n_{F}(E_{{\bf p}})]- n_{F}(E_{{\bf p}})n_{F}(-E_{{\bf p}+{\bf q}})$,
$F^{(\nu)}_{3}({\bf k},{\bf p},{\bf q})=n_{B}(\omega^{(\nu)}_{{\bf
q}+{\bf k}})[n_{F}(-E_{{\bf p}})-n_{F}(E_{{\bf p}+{\bf q}})]+
n_{F}(-E_{{\bf p}})n_{F}(-E_{{\bf p}+{\bf q}})$, $F^{(\nu)}_{4}
({\bf k},{\bf p},{\bf q})=n_{B}(\omega^{(\nu)}_{{\bf q}+{\bf k}})
[n_{F}(-E_{{\bf p}})-n_{F}(E_{{\bf p}+{\bf q}})]-n_{F}(E_{{\bf p}})
n_{F}(E_{{\bf p}+{\bf q}})$, while the charge carrier quasiparticle
coherent weight $Z_{hF}$ and effective charge carrier gap parameter
$\bar{\Delta}_{h}$ are determined by the following two
self-consistent equations \cite{feng},
\begin{widetext}
\begin{subequations}
\begin{eqnarray}
1&=&{1\over N^{3}}\sum_{{\bf k},{\bf p},{\bf q}}[Zt\gamma_{{\bf k}
+{\bf q}}-Zt'\gamma'_{{\bf k}+{\bf q}}]^{2}\gamma^{(d)}_{{\bf k}-
{\bf p}+{\bf q}}\gamma^{(d)}_{{\bf k}}{Z^{2}_{hF}\over 2E_{h{\bf
k}}}{B_{{\bf p}}B_{{\bf q}}\over\omega_{{\bf p}}\omega_{{\bf q}}}
\left({L_{1}({\bf k},{\bf p},{\bf q})\over (\omega_{{\bf p}}
-\omega_{{\bf q}})^{2}-E^{2}_{h{\bf k}}}-{L_{2}({\bf k},{\bf p},
{\bf q})\over (\omega_{{\bf p}}+\omega_{{\bf q}})^{2}-
E^{2}_{h{\bf k}}}\right ) ,  \\
{1\over Z_{hF}}&=&1+{1\over N^{2}}\sum_{{\bf p},{\bf q}}(Zt
\gamma_{{\bf p}+{\bf k}_{0}}-Zt'\gamma'_{{\bf p}+{\bf k}_{0}})^{2}
Z_{hF}{B_{{\bf p}}B_{{\bf q}}\over 4\omega_{{\bf p}}\omega_{{\bf q}
}}\left({R_{1}({\bf p},{\bf q})\over(\omega_{{\bf p}}-\omega_{{\bf
q}}-E_{h{\bf p}-{\bf q}+{\bf k}_{0}})^{2}}+{R_{2}({\bf p},{\bf q})
\over (\omega_{{\bf p}}-\omega_{{\bf q}}+E_{h{\bf p}-
{\bf q}+{\bf k}_{0}})^{2}}\right . \nonumber \\
&+&\left . {R_{3}({\bf p},{\bf q})\over (\omega_{{\bf p}}+
\omega_{{\bf q}}-E_{h{\bf p}-{\bf q}+{\bf k}_{0}})^{2}}+{R_{4}({\bf
p},{\bf q})\over (\omega_{{\bf p}}+\omega_{{\bf q}}+E_{h{\bf p}-{\bf
q}+{\bf k}_{0}} )^{2}} \right ) ,
\end{eqnarray}
\end{subequations}
\end{widetext}
where ${\bf k_{0}}=[\pi,0]$, $L_{1}({\bf k},{\bf p},{\bf q})=
(\omega_{{\bf p}}-\omega_{{\bf q}})[n_{B}(\omega^{(1)}_{{\bf q}})
-n_{B}(\omega^{(1)}_{{\bf p}})+n_{B}(\omega^{(2)}_{{\bf q}})-n_{B}
(\omega^{(2)}_{{\bf p}})][1-2n_{F}(E_{h{\bf k}})]+E_{h{\bf k}}
[n_{B}(\omega^{(1)}_{{\bf p}})n_{B}(-\omega^{(1)}_{{\bf q}})+n_{B}
(\omega^{(1)}_{{\bf q}})n_{B}(-\omega^{(1)}_{{\bf p}})+n_{B}
(\omega^{(2)}_{{\bf p}})n_{B}(-\omega^{(2)}_{{\bf q}})+n_{B}
(\omega^{(2)}_{{\bf q}})n_{B}(-\omega^{(2)}_{{\bf p}})]$,
$L_{2}({\bf k} ,{\bf p},{\bf q})=(\omega_{{\bf p}}+\omega_{{\bf q}})
[n_{B}(-\omega^{(1)}_{{\bf p}})-n_{B}(\omega^{(1)}_{{\bf q}})+
n_{B}(-\omega^{(2)}_{{\bf p}})-n_{B}(\omega^{(2)}_{{\bf q}})]
[1-2n_{F}(E_{h{\bf k}})]+E_{h{\bf k}}[n_{B}(\omega^{(1)}_{{\bf p}})
n_{B}(\omega^{(2)}_{{\bf q}})+n_{B}(-\omega^{(1)}_{{\bf p}})n_{B}
(-\omega^{(2)}_{{\bf q}})+n_{B}(\omega^{(2)}_{{\bf p}})n_{B}
(\omega^{(1)}_{{\bf q}})+n_{B}(-\omega^{(2)}_{{\bf p}})n_{B}
(-\omega^{(1)}_{{\bf q}})]$, $R_{1}({\bf p},{\bf q}) =
n_{F}(E_{h{\bf p}-{\bf q}+{\bf k}_{0}})\{U^{2}_{h{\bf p}-{\bf q}+
{\bf k}_{0}} [n_{B}(\omega^{(1)}_{{\bf q}}) -
n_{B}(\omega^{(1)}_{{\bf p}})]+V^{2}_{h{\bf p}-{\bf q}+{\bf k}_{0}}
[n_{B}(\omega^{(2)}_{{\bf q}})-n_{B}(\omega^{(2)}_{{\bf p}})]\}-
U^{2}_{h{\bf p}-{\bf q}+{\bf k}_{0}}n_{B}(\omega^{(1)}_{{\bf p}})
n_{B}(-\omega^{(1)}_{{\bf q}})-V^{2}_{h{\bf p}-{\bf q}+{\bf k}_{0}}
n_{B}(\omega^{(2)}_{{\bf p}})n_{B}(-\omega^{(2)}_{{\bf q}})$,
$R_{2}({\bf p},{\bf q})=n_{F}(E_{h{\bf p}-{\bf q}+{\bf k}_{0}})
\{U^{2}_{h{\bf p}-{\bf q}+{\bf k}_{0}}[n_{B}(\omega^{(2)}_{{\bf p}}
)-n_{B}(\omega^{(2)}_{{\bf q}})]+V^{2}_{h{\bf p}-{\bf q}+{\bf k}_{0}
}[n_{B}(\omega^{(1)}_{{\bf p}})-n_{B}(\omega^{(1)}_{{\bf q}})]\}-
U^{2}_{h{\bf p}-{\bf q}+{\bf k}_{0}}n_{B}(\omega^{(2)}_{{\bf q}})
n_{B}(-\omega^{(2)}_{{\bf p}})-V^{2}_{h{\bf p}-{\bf q}+{\bf k}_{0}}
n_{B}(\omega^{(1)}_{{\bf q}})n_{B}(-\omega^{(1)}_{{\bf p}})$,
$R_{3}({\bf p},{\bf q})=n_{F} (E_{h{\bf p}-{\bf q}+{\bf k}_{0}})
\{U^{2}_{h{\bf p}-{\bf q}+{\bf k}_{0}}[n_{B}(\omega^{(2)}_{{\bf q}})
-n_{B}(-\omega^{(1)}_{{\bf p}})]+V^{2}_{h{\bf p}-{\bf q}+{\bf k}_{0}
}[n_{B}(\omega^{(1)}_{{\bf q}})-n_{B}(-\omega^{(2)}_{{\bf p}})]\}+
U^{2}_{h{\bf p}-{\bf q}+{\bf k}_{0}}n_{B}(\omega^{(1)}_{{\bf p}})
n_{B}(\omega^{(2)}_{{\bf q}})+V^{2}_{h{\bf p}-{\bf q}+{\bf k}_{0}}
n_{B}(\omega^{(2)}_{{\bf p}})n_{B}(\omega^{(1)}_{{\bf q}})$,
$R_{4}({\bf p},{\bf q})=n_{F}(E_{h{\bf p}-{\bf q}+{\bf k_{0}}})
\{U^{2}_{h{\bf p}-{\bf q}+{\bf k}_{0}}[n_{B}(-\omega^{(1)}_{{\bf q}}
)-n_{B}(\omega^{(2)}_{{\bf p}})]+V^{2}_{h{\bf p}-{\bf q}+{\bf k}_{0}
}[n_{B}(-\omega^{(2)}_{{\bf q}})-n_{B}(\omega^{(1)}_{{\bf p}})]\}+
U^{2}_{h{\bf p}-{\bf q}+{\bf k}_{0}}n_{B}(-\omega^{(2)}_{{\bf p}})
n_{B}(-\omega^{(1)}_{{\bf q}})+V^{2}_{h{\bf p}-{\bf q}+{\bf k}_{0}}
n_{B}(-\omega^{(1)}_{{\bf p}})n_{B}(-\omega^{(2)}_{{\bf q}})$, with
$U^{2}_{h{\bf p}-{\bf q}+{\bf k}_{0}}=(1+\bar{\xi}_{{\bf p}-{\bf
q}+{\bf k}_{0}}/E_{h{\bf p}-{\bf q}+{\bf k}_{0}})/2$, $V^{2}_{h{\bf
p}-{\bf q}+{\bf k}_{0}} = (1- \bar{\xi}_{{\bf p}-{\bf q}+{\bf
k}_{0}}/E_{h{\bf p}-{\bf q}+{\bf k}_{0}})/2$, and $n_{B} (\omega)$
and $n_{F}(\omega)$ are the boson and fermion distribution
functions, respectively. These two equations (7a) and (7b) must be
solved simultaneously with other self-consistent equations, then all
order parameters, decoupling parameter $\alpha$, and chemical
potential $\mu$ are determined by the self-consistent calculation
\cite{feng}. In this sense, our above self-consistent calculation
for the dynamical spin structure factor under a uniform external
magnetic field is controllable without using adjustable parameters,
which also has been confirmed by a similar self-consistent
calculation for the dynamical spin structure factor in the case
without a uniform external magnetic field \cite{feng}.

With the help of the full spin Green's function (3), we can obtain
the dynamical spin structure factor of cuprate superconductors under
a uniform external magnetic field in the SC-state as,
\begin{widetext}
\begin{eqnarray}
S({\bf k},\omega)&=&-2[1+n_{B}(\omega)]{\rm Im}D({\bf k},\omega)
=-{2[1+n_{B}(\omega)]B^{2}_{{\bf k}}{\rm Im}\Sigma^{(s)}({\bf k},
\omega)\over [(\omega-2\varepsilon_{B})^{2}-\omega^{2}_{{\bf k}}-
B_{{\bf k}}{\rm Re}\Sigma^{(s)}({\bf k},\omega)]^{2}+[B_{{\bf k}}
{\rm Im}\Sigma^{(s)}({\bf k},\omega)]^{2}},
\end{eqnarray}
\end{widetext}
where ${\rm Im}\Sigma^{(s)}({\bf k},\omega)$ and ${\rm Re}
\Sigma^{(s)}({\bf k}, \omega)$ are the imaginary and real parts of
the spin self-energy function (6), respectively.

\section{Magnetic field induced incommensurate magnetic resonance}

We are now ready to discuss the influence of a uniform external
magnetic field on the dynamical spin response of cuprate
superconductors in the SC state. For cuprate superconductors, the
commonly used parameters in this paper are chosen as $t/J=2.5$ and
$t'/t=0.3$ with a reasonably estimative value of $J\sim 120$ meV
\cite{shamoto}. At zero external magnetic field ($B=0$), we have
reproduced the previous results \cite{feng}. Furthermore, we have
also performed the calculation for the dynamical spin structure
factor $S({\bf k},\omega)$ in Eq. (8) with a uniform external
magnetic field, and the results of $S({\bf k},\omega)$ in the
($k_{x},k_{y}$) plane for doping $x=0.15$ with temperature
$T=0.002J$ and Zeeman magnetic energy $\varepsilon_{B}=0.01J=1.2$
meV (then the corresponding external magnetic field $B\approx 20$
Tesla) at energy (a) $\omega =0.08J=9.6$ meV, (b) $\omega =0.31J=
37.2$ meV, and (c) $\omega =0.59J=70.8$ meV are plotted in Fig. 1.
In comparison with the previous results without a uniform external
magnetic field \cite{feng}, our present most surprising results
involve the external magnetic field dependence of the resonance
scattering form, i.e., with increasing the external magnetic field
$B$, although the IC magnetic scattering from both low and high
energies is rather robust, the commensurate magnetic resonance
scattering peak is broadened, and is shifted from the AF ordering
wave vector ${\bf Q}$ to the IC magnetic scattering peaks with the
incommensurability $\delta_{r}$. The main difference is that the
resonance response occurs at an IC in the presence of a uniform
external magnetic field, rather than commensurate in the case of
zero external magnetic field. In this sense, we call such magnetic
resonance as the IC magnetic resonance. Experimentally, the growth
of the low energy IC magnetic resonance scattering due to the
presence of an external magnetic field has been observed from the
cuprate superconductor La$_{2-x}$Sr$_{x}$CuO$_{4}$ \cite{tranquada},
which is qualitatively consistent with our theoretical predictions.
For cuprate superconductors, the upper critical magnetic field at
which superconductivity is completely destroyed is $50$ Tesla or
greater around the optimal doping \cite{wang}. Therefore the present
result is remarkable because the magnitude of the applied external
magnetic field is much less than the upper critical magnetic field
of cuprate superconductors. It has been shown that the magnetic
resonance scattering is very sensitive to the SC pairing, and the
external magnetic field induced the IC magnetic resonance scattering
is always accompanied with a breaking of the SC pairing \cite{dai1},
this leads to a reduction of the SC transition temperature in
cuprate superconductors.

\begin{figure}
\includegraphics[scale=0.70]{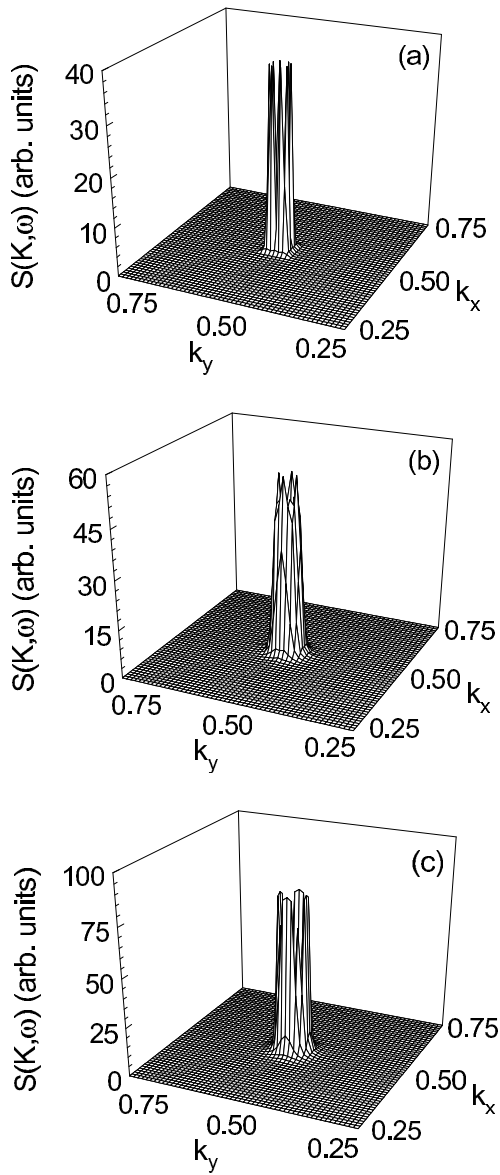}
\caption{The dynamical spin structure factor $S({\bf k},\omega)$ in
the ($k_{x},k_{y}$) plane at $x=0.15$ with $T=0.002J$ and
$\varepsilon_{B}=0.01J$ for $t/J=2.5$ and $t'/t=0.3$ at energy (a)
$\omega =0.08J$, (b) $\omega=0.31J$, and (c) $\omega =0.59J$.}
\end{figure}

\begin{figure}
\includegraphics[scale=0.45]{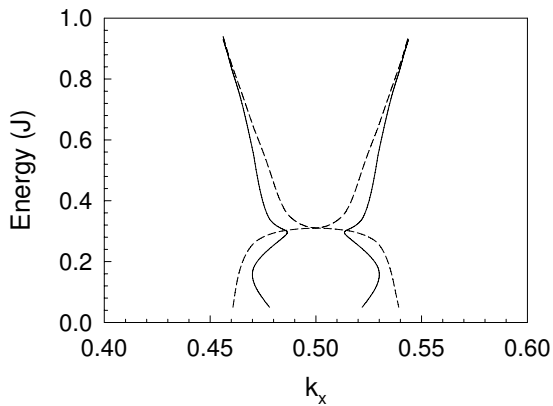}
\caption{The evolution of the magnetic scattering peaks with energy
at $x=0.15$ in $T=0.002J$ for $t/J=2.5$ and $t'/t=0.3$ with
$\varepsilon_{B}=0.01J$ (solid line) and $\varepsilon_{B}=0$ (dashed
line).}
\end{figure}

Having shown the presence of the IC magnetic resonance scattering
under a uniform external magnetic field, it is important to
determine its dispersion as the outcome will allow a direct
comparison of the magnetic excitation spectra with and without a
uniform external magnetic field. In Fig. 2, we plot the evolution of
the magnetic scattering peaks with energy for $x=0.15$ in $T=0.002J$
with $\varepsilon_{B}=0.01J= 1.2$ meV ($B\approx 20$ Tesla) (solid
line). For comparison, the corresponding result for $x=0.15$ in
$T=0.002J$ with the same set of parameters except for
$\varepsilon_{B}=0$ ($B=0$) is also shown in Fig. 2 (dashed line).
As in the previous work \cite{feng}, the dispersion of the magnetic
scattering in the case of zero external magnetic field has an
hourglass shape. However, under a modest external magnetic field
$B\approx 20$ Tesla, although there is no strong external magnetic
field induced change for the IC magnetic scattering at higher energy
$\omega\sim 0.7J$, the magnetic scattering around both intermediate
and low energies is dramatically changed, in qualitative agreement
with the INS experiments \cite{tranquada,bourges1,dai1}. In
particular, although the part above $0.16J\approx 19$ meV seems to
be an hourglass-like dispersion, this hourglass-like dispersion
breaks down at lower energy $\omega< 0.16J \approx 19$ meV. These
are much different from the dispersion in the case of zero external
magnetic field.

\begin{figure}
\includegraphics[scale=0.45]{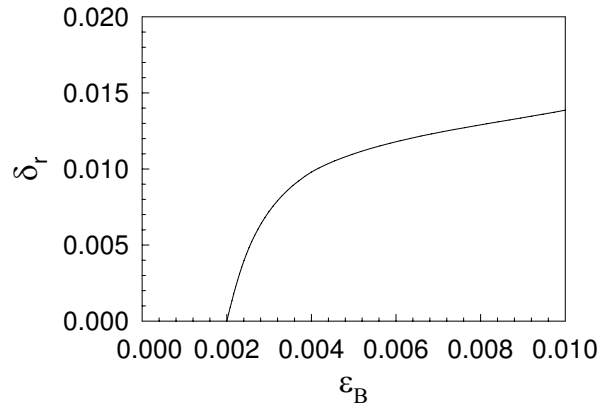}
\caption{The incommensurability of the incommensurate resonance
scattering at $x=0.15$ in $T=0.002J$ for $t/J=2.5$ and $t'/t=0.3$ as
a function of the external magnetic field.}
\end{figure}

\begin{figure}
\includegraphics[scale=0.70]{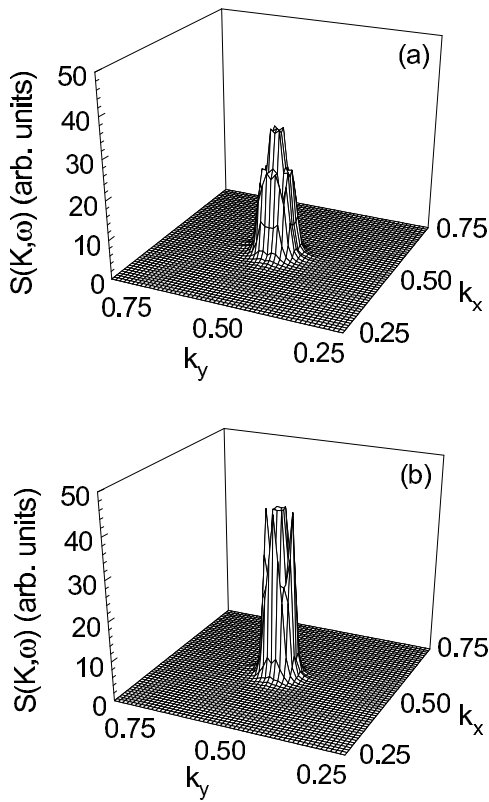}
\caption{The dynamical spin structure factor $S({\bf k},\omega)$ in
the ($k_{x},k_{y}$) plane in $x=0.15$ and $T=0.002J$ for $t/J=2.5$
and $t'/t=0.3$ with (a) $\varepsilon_{B}=0.002J$ and (b)
$\varepsilon_{B}=0.005J$ at $\omega =0.31J$.}
\end{figure}

Now we turn to discuss that how strong external magnetic field can
induce the IC resonance scattering in cuprate superconductors in the
SC state. We have made a series of calculations for the resonance
energy at different external magnetic fields, and the result of the
incommensurability of the IC resonance scattering $\delta_{r}$ for
$x=0.15$ in $T=0.002J$ as a function of a uniform external magnetic
field $B$ is plotted in Fig. 3. Obviously, the incommensurability
$\delta_{r}$ increases with increasing the external magnetic field.
For a better understanding of the influence of a uniform external
magnetic field on the resonance scattering, we plot the dynamical
spin structure factor $S({\bf k},\omega)$ in the ($k_{x},k_{y}$)
plane for $x=0.15$ and $T=0.002J$ with (a) $\varepsilon_{B}=
0.002J=0.24$ meV (then the corresponding external magnetic field
$B\approx 4$ Tesla) and (b) $\varepsilon_{B}= 0.005J=0.6$ meV (then
the corresponding external magnetic field $B\approx 10$ Tesla) at
$\omega =0.31J= 37.2$ meV in Fig. 4. In comparison with Fig. 1(b),
we therefore find that there are two critical values of the Zeeman
magnetic energy $\varepsilon^{(c)}_{B1}\approx 0.002J=0.24$meV (the
corresponding critical external magnetic field $B_{c1}\approx 4$
Tesla) and $\varepsilon^{(c)}_{B2}\approx 0.005J=0.6$meV (the
corresponding critical external magnetic field $B_{c2}\approx 10$
Tesla). When $B>B_{c2}$, the external magnetic field is strong
enough to induce the IC resonance scattering. On the other hand,
when $B_{c1}<B<B_{c2}$, the commensurate resonance scattering peak
is broadened, and remains at the same energy position as the zero
external magnetic field case with a comparable amplitude, which is
furthermore in qualitative agreement with the INS experiments
\cite{bourges1,dai1}.

The physical interpretation to the above obtained results can be
found from the property of the spin excitation spectrum. In contrast
to the case of zero external magnetic field, the MF spin excitation
spectrum has two branches, $\omega^{(1)}_{{\bf k}}=\omega_{{\bf k}}+
2\varepsilon_{B}$ and $\omega^{(2)}_{{\bf k}}=\omega_{{\bf k}}-
2\varepsilon_{B}$, in Eq. (4) under a uniform external magnetic
field as mentioned in Sec. II. Since both MF spin excitation spectra
$\omega^{(1)}_{{\bf k}} $ and $\omega^{(2)}_{{\bf k}}$ and spin
self-energy function $\Sigma^{(s)}({\bf k},\omega)$ in Eq. (6) are
strong external magnetic field dependent, this leads to that the
renormalized spin excitation spectrum $(\Omega_{{\bf k}}-2
\varepsilon_{B})^{2} = \omega^{2}_{{\bf k}}+{\rm Re}\Sigma^{(s)}
({\bf k},\Omega_{{\bf k}})$ in Eqs. (3) and (8) also is strong
external magnetic field dependent. As in the case of zero external
magnetic field \cite{feng}, the dynamical spin structure factor
$S({\bf k},\omega)$ in Eq. (8) under a uniform external magnetic
field has a well-defined resonance character, where $S({\bf
k},\omega)$ exhibits peaks when the incoming neutron energy $\omega$
is equal to the renormalized spin excitation, i.e.,
\begin{eqnarray}
W({\bf k}_{c},\omega)&\equiv& [(\omega-2\varepsilon_{B})^{2}-
\omega_{{\bf k}_{c}}^{2} - B_{{\bf k}_{c}} {\rm Re} \Sigma^{(s)}
({\bf k}_{c},\omega)]^{2}\nonumber \\
&\sim& 0,~~
\end{eqnarray}
for certain critical wave vectors ${\bf k}_{c}={\bf k}^{(L)}_{c}$ at
low energy, ${\bf k}_{c}={\bf k}^{(I)}_{c}$ at intermediate energy,
and ${\bf k}_{c}={\bf k}^{(H)}_{c}$ at high energy, then the weight
of these peaks is dominated by the inverse of the imaginary part of
the spin self-energy $1/{\rm Im} \Sigma^{(s)}({\bf k}^{(L)}_{c},
\omega)$ at low energy, $1/{\rm Im} \Sigma^{(s)}({\bf k}^{(I)}_{c},
\omega)$ at intermediate energy, and $1/{\rm Im}\Sigma^{(s)}({\bf
k}^{(H)}_{c}, \omega)$ at high energy, respectively. In this sense,
the essential physics of the external magnetic field dependence of
the dynamical spin response is almost the same as in the case of
zero magnetic field. However, as seen from Eqs. (6), (8), and (9), a
modest external magnetic field mainly effects the behavior of the
dynamical spin response around low and intermediate energies, and
therefore leads to some changes of the dynamical spin response
around low and intermediate energies. This is followed by a fact
that the magnitude of the applied uniform external magnetic field is
much less than the upper critical magnetic field of cuprate
superconductors, i.e., the Zeeman magnetic energy
$2\varepsilon_{B}/J =0.02\ll 1$ in Eqs. (6), (8), and (9), then the
renormalized spin excitation spectrum at high energy can be reduced
approximately as $(\omega- 2\varepsilon_{B})^{2}=\omega^{2}_{{\bf
k}}+{\rm Re}\Sigma^{(s)}({\bf k},\omega)\approx\omega$ in Eqs. (6),
(8), and (9). This is why there is only a small influence of a
modest external magnetic field on the IC magnetic scattering at
hight energy. However, around low and intermediate energies, this
small Zeeman magnetic energy $\varepsilon_{B}$ in Eqs. (6), (8), and
(9) plays an important role that reduces the range of the IC
magnetic scattering at low energy and splits the commensurate
resonance peak at zero external magnetic field into the IC resonance
peaks, then the IC magnetic resonance scattering appears.
Furthermore, at the heavily low energy regime $\omega\ll 0.16J$, the
magnitude of the Zeeman magnetic energy $2\varepsilon_{B}=0.02J$ is
comparable with these incoming neutron energies, where both incoming
lower neutron energy and Zeeman magnetic energy dominate the IC
magnetic scattering, then the hourglass-like dispersion breaks down.

\section{Summary and discussions}

In summary, we have discussed the influence of a uniform external
magnetic field on the dynamical spin response of cuprate
superconductors in the SC state based on the kinetic energy driven
SC mechanism. Our results show that the magnetic scattering around
low and intermediate energies is dramatically changed with a modest
external magnetic field. With increasing the external magnetic
field, although the IC magnetic scattering from both low and high
energies is rather robust, the commensurate magnetic resonance
scattering peak is broadened \cite{bourges1,dai1}. In particular,
the part above $0.16J\approx 19$ meV seems to be an hourglass-like
dispersion, which breaks down at the heavily low energy regime
$\omega< 0.16J \approx 19$ meV. The theory also predicts that the
commensurate magnetic resonance scattering at zero external magnetic
field is induced into the IC magnetic resonance scattering by
applying a uniform external magnetic field large enough, which
should be verified by further experiments.

From the INS experimental results, it is shown that although some of
the IC magnetic scattering properties have been observed in the
normal state, the magnetic resonance scattering is the main new
feature that appears into the SC state
\cite{tranquada1,dai,yamada,bourges,he,hayden,wilson}. In
particular, applying a uniform external magnetic field large enough
to suppress superconductivity would yield a spectrum identical to
that measured at normal state \cite{ando}. Incorporating these
experimental results, our present result seems to show that the
external magnetic field causes the behavior of the dynamical spin
response to become more like that of the normal state. Moreover, in
our present discussions, the magnitude of an applied external
magnetic field is much less than the upper critical magnetic field
for cuprate superconductors as mentioned above, and therefore we
believe that both commensurate magnetic resonance scattering at zero
external magnetic field and IC magnetic resonance scattering at an
applied modest external magnetic field are universal features of
cuprate superconductors.

\acknowledgments

The authors would like to thank Professor P. Dai for the helpful
discussions. This work was supported by the National Natural Science
Foundation of China under Grant No. 10774015, and the funds from the
Ministry of Science and Technology of China under Grant Nos.
2006CB601002 and 2006CB921300.

\end{document}